\documentclass[epj]{svjour}
\usepackage{graphics}
\input{psfig.sty}
\begin{document}
\title{Renormalization Scale-Fixing  for Complex Scattering Amplitudes}
\author{Stanley J. Brodsky \inst{1} \and Felipe J. Llanes-Estrada
\inst{2}
}                     
\offprints{}          
\institute{Stanford Linear Accelerator Center,  Stanford University, \\ 
Stanford, California, 94309 \hfill\break
\email{sjbth@slac.stanford.edu} \and  Depto. F\'{\i}sica Te\'orica
I, Fac. Cc. F\'{\i}sicas, Universidad Complutense de Madrid, \\ 28040 
Madrid, Spain. \hfill\break \email{fllanes@fis.ucm.es} }
%
%
\abstract{ We show how to fix the renormalization scale for
hard-scattering exclusive processes such as deeply virtual meson
electroproduction by applying the BLM prescription to the imaginary
part of the scattering amplitude  and employing a fixed-t dispersion
relation to obtain the scale-fixed real part. In this way we resolve the
ambiguity in BLM renormalization scale-setting for complex
scattering amplitudes.  We illustrate this by computing  the $H$ generalized parton distribution at leading twist in an analytic
quark-diquark model for the parton-proton scattering
amplitude which can incorporate Regge exchange contributions 
characteristic of the deep inelastic structure functions.
} 
\maketitle
%
\section{BLM renormalization scale setting} \label{BLMsec}

A typical QCD amplitude for an exclusive process can be calculated
as a power series in the strong coupling constant
\begin{equation}
A=A^{(0)} +A^{(1)}\alpha_s(\mu^2)+A^{(2)}\alpha_s^2(\mu^2)+ \cdot \cdot 
\cdot \end{equation}
The renormalization scale $\mu$  of the running coupling in such
processes can be set systematically in QCD without ambiguity at each
order in perturbation theory using the Brodsky-Lepage-Mackenzie
(BLM) 
method~\cite{Brodsky:1982gc,Brodsky:1997dh,Grunberg:1991ac}.

The BLM scale is derived order-by-order by incorporating the
non-conformal terms associated with the $\beta$ function into the
argument of the running coupling.   This can be done systematically
using the skeleton expansion~\cite{Brodsky:2000cr,Brodsky:1998mf}.
The scale determined by the BLM method is consistent with (a) the
transitivity and other properties of the renormalization
group~\cite{Stueckelberg:1953dz} (b) the renormalization group
principle that relations between observables must be independent of
the choice of intermediate renormalization
scheme~\cite{Brodsky:1995wi,Brodsky:1995if}, and (c) the location of
the analytic cut structure of amplitudes at each flavor threshold.
The nonconformal terms involving the  QCD $\beta$ function are all absorbed by
the scale choice. The coefficients of the perturbative series remaining after
BLM-scale-setting are thus the same as those of a conformally
invariant theory with $\beta=0$.    In practice, one can often simply
use the flavor dependence of the series to tag the nonconformal
$\beta$ dependence in perturbation theory; i.e., the BLM procedure
resums the terms involving  $n_f$  associated with the running of
the QCD coupling.

Non-Abelian gauge theory based on $SU(N_C)$ symmetry becomes an
Abelian QED-like theory  in the  limit $N_C \to 0$  while keeping
$\alpha= C_F \alpha_s$ and $n_\ell = {n_{\rm eff }/ 2 C_F}$
fixed~\cite{Brodsky:1997jk}.    Here $C_F = {(N^2_C-1)/ 2 N_C}$.
The BLM scale reduces properly to the standard QED scale in this
analytic  limit. For example, consider the vacuum polarization
lepton-loop correction to $e^+ e^- \to e^+ e^-$ in QED. The
amplitude must be proportional to $\alpha(s)$ since this gives the
correct cut of the forward amplitude at the lepton pair threshold $s
= 4 m^2_\ell.$  Thus the renormalization scale $\mu^2_R = s$ is
exact and unambiguous in the  conventional QED Goldberger-Low
scheme~\cite{Gell-Mann:1954fq}.  If one chooses any other scale
$\mu^2_R \ne  s$, the scale  $\mu^2_R = s$ will be restored when one
sums all bubble graphs. The BLM procedure is thus consistent with
the Abelian limit and the proper cut structure of amplitudes.

\section{Difficulties in using the mean value theorem to set the
BLM Scale}

The BLM scale at leading order has a simple physical interpretation:
it is identical to the photon virtuality in QED applications and the
mean gluon virtuality  in QCD when one uses  physical schemes which
generalize the QED scheme such as the pinch
scheme~\cite{Brodsky:1997dh,Cornwall:1981zr} and the
$\alpha_V$~\cite{Brodsky:1998mf,Appelquist:1977es} scheme 
defined from the QCD static potential.  The number of flavors active in
virtual corrections to a given process is evident from the BLM scale
choice: the BLM method  sets the renormalization scale so that
flavor number is changed properly in any renormalization scheme, including  the
${\overline{MS}}$ scheme~\cite{Brodsky:1997dh,Brodsky:1999fr}.

In effect, the  BLM prescription identifies the renormalization scale and the gluon virtuality  by
eliminating the dependence on the number of flavors from the
$O(\alpha_s)$ (expanding $\alpha_s$ itself) and $O(\alpha_s^2)$ nonconformal
terms in  the perturbative amplitude. Typically, a QCD amplitude
involves an integral over the momentum running through the  gluon
propagator. Therefore the argument of  the coupling, if taken
to be the momentum flowing through the gluon, varies
through the integration phase space. A mean $\bar{Q^2}$ can be
extracted from the integral if the mean value theorem (MVT) of
integral calculus can be applied. The essential requirement for the
applicability of the MVT is that the function  being evaluated at
its mean value has to be continuous through the interval and the
weight function has to be Riemann-integrable; 
this includes weight functions bounded and continuous in the 
range of integration. However, these necessary conditions are 
not a property of a principal value integrand associated with a 
pole which  appears, for example, in Compton scattering and 
deeply virtual meson electroproduction (DVME); as we show 
explicitly in Section \ref{failure} below, the MVT  does not 
apply for these amplitudes.

It has been recently pointed out in Ref. \cite{Anikin:2004jb} that the  MVT
prescription for determining the BLM procedure fails for amplitudes which are genuinely complex, that
is, display non-vanishing real and imaginary parts. The authors then
argue that there is no guarantee that the BLM prescription will
yield the same answer for both parts of the amplitude. Worse, in the
particular example that they study, $\rho$ and $\widehat{\pi}_1$
electroproduction, the scale obtained from the real part becomes
discontinuous (zero to infinity) at a particular kinematical point
due to a divergence in some of the intermediate functions.

In this paper we note that in a quantum field theory, the real and
imaginary parts of a scattering amplitude are not independent but
are constrained due to causality, locality, and Lorentz
invariance. This manifests itself in the form of the dispersion
relations traditionally used in meson photoproduction
\cite{Chew:1957tf,oehme} to link both parts of the amplitude. By
examining a simple example in the next subsection, we show that the
correct prescription for finding the BLM renormalization scales
is to first fix the scale in the imaginary part, and then
subsequently, the real part can be obtained by means of a dispersion
relation.   The BLM scales for the real and imaginary parts are 
thus not generally equal.  We perform an  explicit calculation 
for  longitudinally polarized vector meson  electroproduction at  
nonzero skewness in Section \ref{meson}.

\subsection{Example of the failure of the Mean Value Theorem}
\label{failure}

To illustrate how the MVT can fail for an amplitudes which contains a pole, consider the following simple integral:
\begin{equation}
I[f](y)=\int_{-1}^{1} f(x) \frac{dx}{x-y+i\epsilon} \
\end{equation}
which is a functional of $f$ and a function of $y$. The MVT would state 
that there would exist a certain
$\bar{x}(y)$ in the interval $(-1,1)$  such that
\begin{equation}
I[f](y) = f(\bar{x}(y)) I[1](y) \ .
\end{equation}
For $f={\rm constant}$ the theorem holds trivially. But now
consider a linear function $f(x)=Ax+B$, with $A,B$ arbitrary real
constants. Substituting Cauchy's principal value for distributions
(the imaginary part arising for $y\in(-1,1)$)
\begin{equation}
\frac{1}{x-y+i\epsilon} = PV\left[ \frac{1}{x-y} \right]
-i\pi \delta(x-y),
\end{equation}
one easily finds
\begin{eqnarray}
I[f](y)&=& 2A + (B+Ay)\left( \log \left( \frac{1-y}{1+y}\right)-i\pi
\right) \\ \nonumber &=& 2A + f(y) I[1](y) \ .
\end{eqnarray}
The MVT thus holds for the imaginary part, but it fails for
the real part by the term $2A$. Even in the instance where $f$
could be chosen so that a certain $\bar{x}$ would satisfy the
theorem for the real part, there would be no reason for $\bar{x}$
to be the same for the imaginary part. 

In this example, the function $I[f](y)$ is cut in the complex $y$
plane in the interval $[-1,1]$ of the real axis.   The function satisfies an unsubtracted
dispersion relation:
\begin{eqnarray} \label{simpledisp}
I[f](y) &=& \frac{1}{2\pi i} \int_{\rm cut} \frac{dy'
I[f](y')}{y-y'-i\epsilon}\\ \nonumber Re I[f](y) &=& \frac{1}{2}
\int {\rm PV}\left[ \frac{1}{y-y'} \right] dy' Im I[f](y') ,
\end{eqnarray}
and thus the discontinuity across the cut is sufficient to
reconstruct the whole function. 
In this trivial example the
discontinuity is
\begin{equation}
I[f](y+i\epsilon)-I[f](y-i\epsilon)=-2\pi i f(y)
\end{equation}
and the  mean value ``scale'' for the imaginary part is $\bar{x}(y)=y$.
Once this $\bar{x}$ has been  chosen, the full amplitude is easily
reconstructed from the dispersion relation.

This procedure is not restricted to linear functions.  For an arbitrary
polynomial one can expand
around $y$,
\begin{equation}
f(x)=f(y) + \sum_{n=1}^{N} \frac{f^{(n)}(y)}{n!} (x-y)^n
\end{equation}
we have now
\begin{equation}
I[f](y)= f(y) I[1](y)+ \sum_{n=1}^N \frac{f^{(n)}(y)}{n\cdot n!}
( (1-y)^n - (-1-y)^n)\ .
\end{equation}
Again the second term is ``unexpected'' and a naive application of
the mean value theorem to the real part fails. This additional  term is a sum of
binomials of increasing degree, which in the limit $N\to \infty$,
can be used to construct any entire function. Since it is analytic, it provides no
contribution to the discontinuity across the cut, and the dispersion
relation in Eq. (\ref{simpledisp}) still holds. 

\subsection{Meson electroproduction}\label{meson}

We now turn to the problem of determining the renormalization
scale for vector meson electroproduction $\gamma^* p \to V^0 p^\prime$, the critical example studied  in
Ref. \cite{Anikin:2004jb}.
We will employ the same kinematics, frame choice, and conventions as in
Ref. \cite{Brodsky:2000xy}, It will be  useful to note here that in the
asymmetric frame employed, the skewness variable plays the role of
Bjorken's  $x$, and
\begin{equation} \label{skewnessdef}
\zeta = \frac{Q^2}{Q^2+s-M_N^2} ;\ Q^2= s+t+u-2M_N^2 \ .
\end{equation}
The skewness is in the interval
\begin{equation}
\zeta \in \left[0  , \frac{-t}{2M^2} \left( \sqrt{1+ \frac{4M^2}{(-t)}} -1
\right) \right] \ .
\end{equation}

As a specific example of meson electroproduction we
consider the production of a $\rho^0$ meson by  a
longitudinally polarized virtual photon $\gamma^*_L$ with large virtuality
$Q^2  = -q^2$~\cite{Brodsky:1994kf,Collins:1996fb}. 
The relation of the differential 
cross section to the generalized parton distributions is well known
\cite{Vanderhaeghen:1999xj}
\begin{eqnarray} \label{rhocross}
\frac{d\sigma_L}{dt} = \frac{1}{16\pi(s-M_N^2)\Lambda^{1/2}
(s,-Q^2,M_N^2)}\frac{1}{2}\sum_{\lambda\lambda'}
\arrowvert M^L\arrowvert^2
\end{eqnarray}
where $\Lambda^{1/2}$ is the  K\"allen's function,
\begin{equation}
\Lambda^{1/2}(a,b,c)=\sqrt{a^2+b^2+c^2-2(ab+ac+bc)}
\end{equation}
and the amplitude to leading order in $\alpha_s$  and  $Q$, 
ignoring the $E$ Generalized Parton Distribution (GPD), is
\begin{eqnarray} \label{rhoamp}
iM^L_{\rho^0} &=& -ie \frac{4}{9}\ \frac{1}{Q}\ 4\pi \alpha_s(\mu^2)
\bar{U}_\lambda(P) \gamma^+ U_{\lambda'}(P')\nonumber \\
&&\times \int_0^1dz \Phi_\rho(z)\frac{1}{4zP^+}  \int_{\zeta-1}^1 \!
\! dx
H^P_{\rho^0_L}(x,\zeta,t)\\
&\quad&\times  \left[ \frac{1}{x-\zeta-i\epsilon}+
\frac{1}{x+i\epsilon} \right] .\nonumber
\end{eqnarray}
We use a model for the GPD $H$ in this expression specified below in 
section  \ref{Hmodeling}. Much useful information on GPD's has been 
collected in \cite{Diehl:2003ny}. Notice that at large $s$ one 
should also take into account double gluon exchange between the nucleon 
and the photon projectile (exiting the reaction as a meson).
Substituting now
\begin{equation}
\bar{U}_\lambda(P)\gamma^+ U_{\lambda'}(P') = 2P^+
\delta_{\lambda\lambda'} \frac{\sqrt{1-\zeta}}{1-\zeta/2}
\end{equation}
and employing the asymptotic value for the vector meson distribution
amplitude~\cite{Lepage:1980fj}
\begin{equation}  \label{mesonDA}
\Phi_\rho(z)=6z(1-z)f_\rho
\end{equation}
with $f_\rho=0.216\ GeV$
we obtain
\begin{eqnarray}\label{immeson}
{\rm Im}\ M^L_{\rho^0}=  \frac{-e\sqrt{1-\zeta}}{1-\zeta/2}
\frac{4}{9}\frac{3f_\rho}{2} 4\pi^2\alpha_s(\mu^2_{BLM})
\delta_{\lambda\lambda'}\frac{1}{Q} \\ \nonumber
\left[
H^P_{\rho^0_L}(0,\zeta,t)-H^P_{\rho^0_L}(\zeta,\zeta,t) \right] \ .
\end{eqnarray}

Once the BLM scale is fixed, as in Eq. (\ref{BLM}) below, the real
part of the amplitude to leading order can be accessed by a
dispersion relation
\begin{equation} \label{remeson}
{\rm Re} \ M^L_{\rho^0}= \frac{1}{\pi}\int_{(M_N+m_\rho)^2}^{\infty}
{\rm Im}\ M^L_{\rho^0} {\rm PV}\left[ \frac{1}{s'-s} \right] ds'\ .
\end{equation}
A few remarks are in order: the $\frac{1}{Q}$ factor in Eq.
(\ref{immeson}) makes the dispersion integral in Eq. (\ref{remeson})
convergent at fixed $\zeta$ (the Bjorken limit). This is the
kinematical case traditionally considered in meson electroproduction
\cite{Vanderhaeghen:1999xj}. Further, the threshold for the integral
is irrelevant if one considers the large $Q$ behavior of
$M^L_{\rho^0}$. 
The use of a dispersion relation at fixed positive skewness $\zeta$ 
induces a left cut in the complex $s$ plane due to the photon becoming 
timelike there. This and the usual $t$-channel left cut
are consistently neglected in eq. \ref{remeson} as they are 
$1/s$ ($1/Q^2$ at fixed $\zeta$) suppressed.
At last and most important, the
$1/Q$ factor in  Eq. (\ref{remeson}) relative to the corresponding
Compton amplitude, makes the
Compton cross section fall much slower in $Q$ than its meson 
electroproduction counterpart.
This is due to the extra gluon necessary to construct the LO 
distribution amplitude of the emitted meson 
and is a striking manifestation of the $J=0$ fixed pole which appears in 
the Compton amplitude due to the quasi-local coupling of two currents 
to the propagating quark.~ \cite{inpreparation}

Let us now turn to setting the BLM renormalization scale for meson electroproduction. 
We shall disregard the
$\mu$ evolution of $H$ as an irrelevant complication for our
application. Since
the leading order is $\alpha_s$, and the NLO amplitude is available,
we can follow \cite{Anikin:2004jb}.  One identifies the NLO contributions from the vacuum
polarization (which can be picked up from its dependence on $N_f$)
and imposes the BLM scale-fixing condition on the imaginary part of the
amplitude
\begin{equation}
{\rm Im} \int_0^1 dz \Phi_{\rho}(z)\int_{\zeta-1}^1 \frac{dx}{1-\zeta/2}
H^P_{\rho^0_L}(x,\zeta,t)\cdot T =0
\end{equation}
with the $\overline{MS}$ expression
\begin{eqnarray}
T &=&  \frac{1}{z(x-i\epsilon)}\left[ \frac{5}{3} -\log\left(
\frac{zx}{\zeta}\right) -\log\left( \frac{Q^2}{\mu^2}\right) \right]
\\ && +\frac{1}{(1-z)} \frac{1}{x-\zeta+i\epsilon} \bigg[
\frac{5}{3} -\log\left( \frac{(1-z)(x-\zeta)}{\zeta} \right)\nonumber\\
&& -\log\left(\frac{Q^2}{\mu^2}\right) \bigg] \ . \nonumber
\end{eqnarray}
Following \cite{Belitsky:2001nq} we first notice that the  $z$
integrals are trivial  when employing the meson distribution amplitude in
Eq. (\ref{mesonDA}),
\begin{eqnarray}
\int_0^1 dz \Phi_\rho (z)\frac{\log z}{z}&=& -\frac{9f_\rho}{2} \nonumber \\
\int_0^1 dz \Phi_\rho (z)\frac{\log (1-z)}{1-z}&=&
-\frac{9f_\rho}{2} \\ \int_0^1 dz \Phi_\rho (z)\frac{1}{z}&=& 3f_\rho \nonumber \\
\int_0^1 dz \Phi_\rho (z)\frac{1}{1-z}&=& 3f_\rho \nonumber
\end{eqnarray}
which allows us to find simpler expressions than those in
\cite{Anikin:2004jb}.
Solving for $\mu^2$ yields the BLM scale
\begin{equation} \label{BLM}
\mu^2 = Q^2 e^{-A_1/A_2}
\end{equation}
in the $\overline{MS}$ scheme where
\begin{eqnarray} \label{fullscalefixing}
A_2&=&\pi
\left[ H(0,\zeta,t)- H(\zeta,\zeta,t) \right]\nonumber \\
A_1 &=& \left(\frac{19}{6}+\log \zeta  \right)A_2 -\pi \left(
-H(\zeta,\zeta,t) \log (1-\zeta) \right.\nonumber \\ && +
\int_{\zeta-1}^0 dx  \frac{H(x,\zeta,t)-H(0,\zeta,t)}{x}
  \\
&& \left.-\int_\zeta^1 dx \frac{H(x,\zeta,t)-H(\zeta,\zeta,t)}{x-\zeta}
\right) \ . \nonumber
\end{eqnarray}
We also note 
that if Eq. (\ref{immeson}) is $N_f$ independent then
Eq. (\ref{remeson}) is also $N_f$ independent, and the goal of the
BLM scale fixing procedure is achieved simultaneously for the real
and imaginary parts of the amplitude thanks to the dispersion
relation. 

We thus have a totally consistent scale setting procedure
at the level of the amplitude.
Choosing to fix the BLM scale at the level of the total
cross section is therefore an unnecessary complication. The reader
may observe that since both $A_1$ and $A_2$ are linear in $H$, the
scale in Eq. (\ref{BLM}) is independent of the absolute
normalization of $H$. This would change if the $E$ or other GPD's
were included in the analysis. The same observation applies to the normalization
of the $\rho$ meson distribution amplitude;  i.e., $\mu$ is
independent of $f_\rho$ in our approximation. Thus the BLM renormalization scales are
effectively independent of the flavor of the meson produced.

\section{Covariant quark-diquark model for the H generalized parton distribution}
\label{Hmodeling}

To complete the evaluation of this specific example, we need a
physically inspired model for $H$. We recall the 
analytic model for virtual Compton scattering introduced in Ref. \cite{Brodsky:1973hm} which is the gauge invariant leading twist extension of  
the ``covariant parton model" for structure functions given in  Ref.  \cite{Landshoff:1970ff}.
This model can also incorporate not only Reggeon exchange in the $t$ channel, but also the $q^2-$ and $t$-independent $J=0$ fixed-pole contribution from the local  coupling of the two photons to the struck quark.  The $J= \alpha_R=1/2$ Regge contributions to the structure functions are evidenced by the $\sqrt x$ behavior of the nonsinglet structure function $F^p_2(x,Q^2)-F^n_2(x,Q^2)$ at small $x$.
Since this model has not been widely used in the
past few years, we find worth readdressing it in the modern context
as a contribution to the current discussion on GPD's. 

The central ansatz of the covariant parton model\cite{Landshoff:1970ff} is to
construct a model of the quark-parton scattering amplitude, a vertex
with two quark legs of momentum $k$, $k'$ and spin indices $i$,$i'$
and two proton legs with momentum $p$, $p'$ and helicities
$\lambda$, $\lambda'$. Since the parton legs are not on-shell, this
amplitude is a function of four different Lorentz scalars, that can
be chosen as the three Mandelstam invariants $s_{pp}=(p+k')^2$,
$t_{pp}=(k'-k)^2=(p'-p)^2$, $u_{pp}=(p-k)^2$ and $k^2$. The squared
momentum of the returning parton can be expressed as
$k'^2=s_{pp}+t_{pp}+u_{pp}-2M_N^2-k^2$. We will denote this amplitude by
\begin{equation}
T_{\lambda \lambda'; i'i}[s_{pp},t_{pp},u_{pp},k^2] \ .
\end{equation}
In terms of the DVME scattering kinematical variables $x$, ${\bf
\Delta}_\perp$, $\zeta$, $t$, and the integration variable $k$, we
have the following expressions for the parton-proton scattering
matrix Lorentz invariants:
\begin{eqnarray} \label{partonprotonkin}
s_{pp}&=&
(1+x-\zeta)P^+
\left( k^- + \frac{M^2+{\bf \Delta}_\perp^2}{(1-\zeta)P^+}  \right)
-({\bf k}_\perp-{\bf \Delta}_\perp)^2\nonumber \\
u_{pp}&=& (P^+k^- -M_N^2)(x-1)-{\bf k}_\perp^2 \nonumber \\
t_{pp}&=& t \\
k^2&=& x P^+ k^- - {\bf k}_\perp^2\nonumber \\
k^{'2}&=& t + M_N^2(1-x)+P^+k^-(x-\zeta)+(M_N^2+{\bf
\Delta}_\perp^2)\nonumber \\
&& \times \left( 1+\frac{x}{1-\zeta}\right) - ({\bf k}_\perp-{\bf
\Delta}_\perp)^2  \nonumber
\end{eqnarray}
(hereafter the subindex $ _{pp}$ can be dropped without confusion).
\begin{figure}
 \psfig{figure=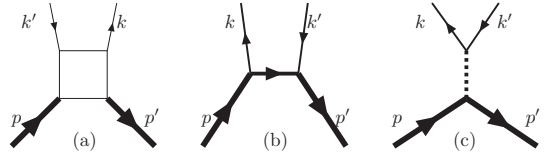,height=2in}
\caption[*]{\label{partonproton} (a) Quark-proton scattering matrix.
(b) Quark-proton u-channel diquark exchange. (c) Quark-proton
$t$-channel pomeron/Reggeon exchange. 
The need for modeling Regge trajectories is avoided by requiring $t\ll 
-1\ GeV^2$, where $\alpha(t)\to -1$\cite{Coon:1974wh} and 
$u^{\alpha(t)}\propto \frac{1}{u}$, which can be reabsorbed in the quark-diquark model.}
\end{figure}
To connect with the modern formulation of deeply virtual Compton scattering in terms of GPD's, let us
note \cite{Diehl:2003ny} that
\begin{eqnarray}
&&\frac{1}{2P^+} \bar{U}(P')_{\lambda'} \left[
H(x,\zeta,t)\gamma^+ - E(x,\zeta,t) \frac{i\sigma^{+\alpha}
\Delta_\alpha}{2M_N} \right] U(P)_\lambda
  \nonumber \\
&&=\int \frac{dy^-}{8\pi} e^{ixP^+y^-/2}
\langle P',\lambda' \arrowvert \bar{\Psi}(0) \gamma^+ \Psi(y)
\arrowvert P,\lambda
\rangle\arrowvert_{y_+=0,\ y_\perp=0}
\nonumber \\
&&=\frac{1}{2}\int \frac{d^4k}{(2\pi)^4} \delta(xP^+-k^+)
\int \frac{d^4k'}{(2\pi)^4} \delta((x-\zeta)P^+-k^{'+})
 \nonumber\\
&&\quad \times \gamma^+_{ii'}
(2\pi)^4  \delta^{(4)}(p+k'-p'+k) T_{\lambda \lambda'; i'i}[s,t,u,k^2]
\end{eqnarray}
where the Fourier modes of the fermion field are
\begin{equation}
\widehat{\Psi}(k)=\int d^4y e^{iky} \Psi(y) \delta^{(2)}({\bf
y}_\perp) \delta(y^+)\ .
\end{equation}
This allows one to construct the leading twist handbag contributions to the GPD's given a model for the
parton-proton scattering amplitude, by integrating over parton
transverse and $(-)$ momentum, and contracting the spins with a
$\gamma^+$ matrix in the parton Dirac space.

The simplest model for the parton-proton amplitude is a tree
level diagram based on perturbation theory with a
proton-quark-diquark vertex (see Fig. \ref{partonproton}). Since we are only 
interested in leading twist effects in the electroproduction
amplitude, we will ignore the axial vector diquark and the
spin dependence of the vertex. The proton spin in this model, with
an $s$-wave proton-quark-diquark vertex, is carried by the struck
quark, so that
\begin{equation}
T_{\lambda \lambda'; i'i}[s,t,u,k^2] = T[s,t,u,k^2] \delta_{\lambda
\lambda'} \delta_{ii'}
\end{equation}
for the $H$ GPD part, that yields an oversimplified yet practical
\begin{equation} \label{GPDpartonprotonrep}
\frac{1}{2P^+} H(x,\zeta,t) = \frac{1}{2} \int \frac{d^4k}{(2\pi)^4}
\delta(P^+x-k^+) T[s,t,u,k^2] \ .
\end{equation}

The amplitude in covariant perturbation theory is
\begin{equation} \label{simplediquark}
i T(s,t,u,k^2,k^{'2})=(ig(k^2))
\frac{i}{(p-k)^2-\lambda^2+i\epsilon}
(ig(k^{'2}))
\end{equation}
Here $\lambda$ plays the role of a scalar diquark mass and $g$ is
the amplitude for the proton-quark-diquark coupling. The stability
condition dictates that the sum of the diquark and parton masses has
to be larger than the proton mass to prevent  the decay to a 
free quark and diquark pair. This feature is not apparent from
Dyson-Schwinger models of the proton-quark-diquark vertex
\cite{Maris:2004ig}, \cite{Bloch:1999ke}, \cite{Oettel:2000jj},
where the diquark mass is of order 800 MeV and decreases
simultaneously with the quark mass at large momentum, but the
Euclidean space formulation does not accommodate the decay. Except
for this point, the work of these authors can perfectly be reused
here as model input.

The vertex functions $g(k^2)$ are known from these
fundamental studies to fall with $k^2$. One could guess
\begin{equation} \label{naivevertex}
g(k^2)=\frac{g}{(k^2-\Lambda^2)+i\epsilon} 
\end{equation}
for $\Lambda$ some constituent or running quark mass (that we will
take fixed here at 0.5 $GeV$, with $g=55$ and $\Lambda=1.6\ GeV$).
This is simply suggested by the observation that $T$ is not a
truncated amplitude at the quark legs, and therefore it 
includes  their propagators. But since all the recoil $t$ dependence
in the form factor calculation needs to be present in the vertex
$g(k^{'2})$ and we know by the QCD counting rules that at large
momentum transfer $F_1(t)\propto t^{-2}$, the simplest workable
model is
\begin{equation} \label{squaredvertex}
g(k^2)=\frac{g}{\left((k^2-\Lambda^2)+i\epsilon\right)^2} \ .
\end{equation}
This can be viewed as a derivative with respect to $\Lambda$ of Eq.
(\ref{naivevertex}); since it can be pulled out of the $k$
integration as a parametric derivative, it  does not alter the 
analytical structure of the DVME amplitude nor the placement of poles of 
$T$ in the $k^-$ plane which  we now study.

\begin{figure}
\psfig{figure=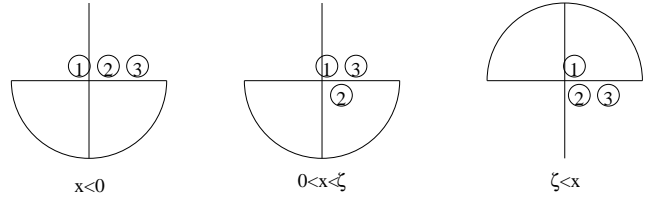,height=1in}
\caption[*]{\label{poleplacement} Integration over the $\kappa^-=
k^-P^+$ variable in the complex plane for the diquark model. From
the figure, $H(x<0)=0$.}
\end{figure}

Consider therefore the model defined by Eq. (\ref{simplediquark})
and (\ref{squaredvertex}). We employ the variable $\kappa^- =
P^+k^-$ so that an inverse power of $P^+$ comes out of the $k^-$
integral and yields
\begin{equation} \label{GPDpartonprotonrep2}
H(x,\zeta,t) = \int_0^\infty \frac{\arrowvert {\bf k}_\perp \arrowvert
d\arrowvert {\bf k}_\perp \arrowvert}{(2\pi)^4} \int_0^{2\pi}d\phi_\perp
\int_0^\infty d\kappa^-  T[s,t,u,k^2] \ .
\end{equation}
in place of Eq. (\ref{GPDpartonprotonrep})  and with $k^+$, $k'$
fixed as above. The parton-proton amplitude is at this order an
holomorphic function of $\kappa^-$ and has three poles, whose
positions are depicted in Fig.~\ref{poleplacement}.
The diquark propagator yields a simple pole denoted as (1) and
given by the condition $u-\lambda^2+i\epsilon=0$. The vertex functions
yield two double poles respectively, denoted by (2) for $k^2-\Lambda^2
+i\epsilon=0$ and (3) for $k^{'2}-\Lambda^2+i\epsilon=0$. These poles are
respectively located at
\begin{eqnarray}
\kappa^-_1 &=& - \frac{\lambda^2+{\bf k}_\perp^2-M_N^2(1-x)}{1-x}
-i\epsilon
\nonumber \\
\kappa^-_2 &=&\frac{\Lambda^2+{\bf k}_\perp^2+i\epsilon}{x}
\\
\kappa^-_3 &=&\frac{\Lambda^2+c^2+i\epsilon}{x-\zeta}\nonumber
\end{eqnarray}
where we have grouped
\begin{eqnarray}
c^2&=& -\bigg[ t+M_N^2(1-x) +  (M_N^2+{\bf \Delta}_\perp^2)\nonumber
\\
&&\times \left( 1+\frac{x}{1-\zeta} \right) -({\bf k}_\perp -{\bf
\Delta}_\perp)^2 \bigg] \ .
\end{eqnarray}

\begin{figure}
\psfig{figure=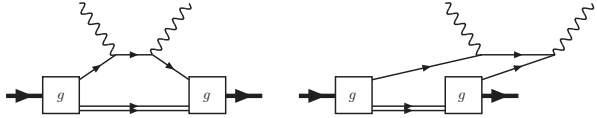,height=1.23in}
\caption[*]{\label{diquarkmodel} Two light-cone time-ordered diagrams for the
handbag diagram with a covariant parton-proton scattering vertex
based on a parton-diquark-proton vertex and second order
perturbation theory. Notice that the second diagram, present in a
covariant model, gives rise to a non-vanishing $H(x\in
(0,\zeta),\zeta,t)$ and  to a well-defined imaginary part, by
providing a finite part from the left needed for continuity at
$x=\zeta$.}
\end{figure}

The $\kappa^-$ integral can now be immediately performed picking up
the residue of the simple or double pole as in
Fig.~\ref{diquarkmodel}. We find
\begin{eqnarray} \label{Hdiquark1}
&&H(x\in(0,\zeta),\zeta,t) = g^2\Lambda^8 \frac{1-\zeta/2}{(2\pi)^3}
\int_0^\infty \arrowvert {\bf k}_\perp\arrowvert d\arrowvert {\bf
k}_\perp\arrowvert
\\
&&\times \int_0^{2\pi} d\phi_\perp \left(
\frac{1}{u-\lambda^2}\frac{1}{(k^{'2}-\Lambda^2)^2}\right) \left(
\frac{x-1}{u-\lambda^2}  + \frac{2(x-\zeta)}{k^{'2}-\Lambda^2}
\right)\nonumber
\end{eqnarray}
evaluated at the pole $k^-=k^-_2$. The second term, being
proportional to $x-\zeta$, does not give a contribution to the imaginary
part of the electroproduction amplitude at leading $Q^2$. We also have
\newpage
\begin{eqnarray} \label{Hdiquark2}
H(x\in(\zeta,1),\zeta,t) = g^2\Lambda^8 \frac{1-\zeta/2}{(2\pi)^3}
\int_0^\infty
\arrowvert {\bf k}_\perp\arrowvert d\arrowvert {\bf k}_\perp\arrowvert
\\ \nonumber
\int_0^{2\pi} d\phi_\perp
\left(\frac{1}{(k^{'2}-\Lambda^2)^2}\frac{1}{(k^2-\Lambda^2)^2}
\right)\
\end{eqnarray}
which gives a positive definite $H$ function.

\section{Numerical results}

\begin{figure}
\psfig{figure=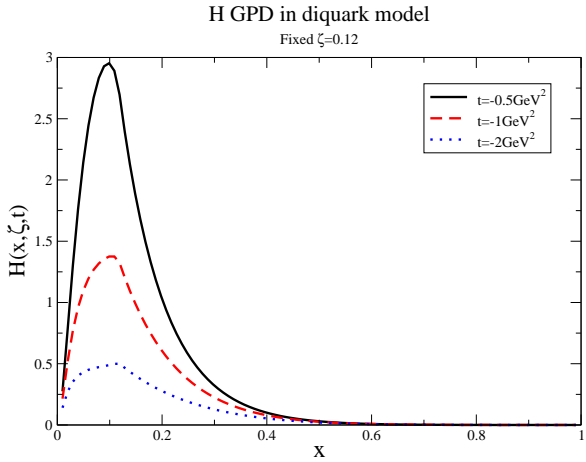,height=2.75in,angle=-90}
\caption[*]{\label{diquarkHoft}  The $t$ dependence of the $H(x,\zeta,t)$
Generalized Parton Distribution in the perturbative diquark model at
fixed $\zeta=0.12$.  Since this model does not have a sea wavefunction
component, it does not give access to the region $\zeta-1<x<0$.
However it is sufficient for the purpose of illustrating BLM scale
fixing via a dispersion relation.}
\end{figure}

\begin{figure}
\psfig{figure=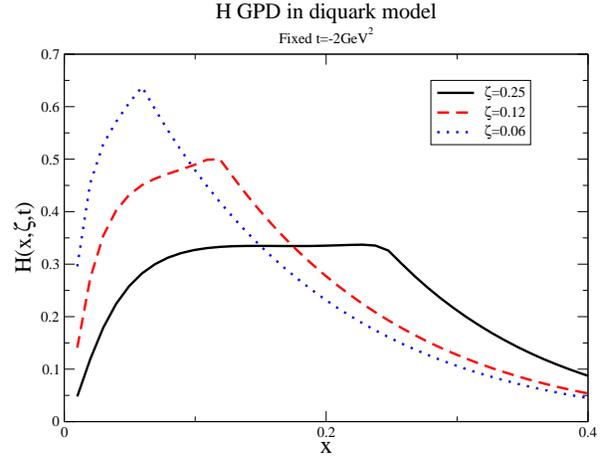,height=2.75in,angle=-90}
\caption[*]{\label{diquarkHofskew} $\zeta$ dependence of the
$H(x,\zeta,t)$ Generalized Parton Distribution in the perturbative
diquark model at fixed $t=-2GeV^2$. Observe that by performing
different pole integrals for $x>\zeta$ and $x<\zeta$, obtaining
continuity at $x=\zeta$ is a non-trivial test of the model and
computations.}
\end{figure}

We now provide  a numerical computation of the above model in order  to illustrate the
approach. The $t$ dependence of the $H$
GPD in this quark-diquark model is plotted in Fig.~\ref{diquarkHoft} .
As can be seen, $H$ falls rapidly with $t$ at fixed $\zeta$. The
region $\zeta-1<x<0$ cannot be accessed with this ``valence'' model.
However, the fact that the vertex employed is covariant gives a
feature absent in the constituent valence light-front quark model,
that is, a non vanishing $H(x=\zeta,\zeta,t)$ and continuity at
$x=\zeta$. This allows for an imaginary part in the
electroproduction amplitude in this minimal model. In
Fig.~\ref{diquarkHofskew} we plot the $\zeta$ dependence  of $H$.
The sharpness of the peak in this graph is controlled by $\lambda$, the
diquark mass.
The diquark model satisfies the simplifying property
\begin{equation}
H(x<0,\zeta,t)=0.
\end{equation}
We then can reproduce the result of Belitsky and Muller for the BLM
scale for the imaginary part of the amplitude;  in our
asymmetric frame Eq. (\ref{fullscalefixing})  their result is
\begin{eqnarray} \label{quotientscale}
\frac{A_1}{A_2}= \frac{19}{6}+\log \left( \frac{\zeta}{1-\zeta}
\right) \\ \nonumber
-\int_\zeta^1 dx \frac{\frac{H(x,\zeta,t)}{H(\zeta,\zeta,t)}-1}
{x-\zeta}
\end{eqnarray}

\begin{figure}
\psfig{figure=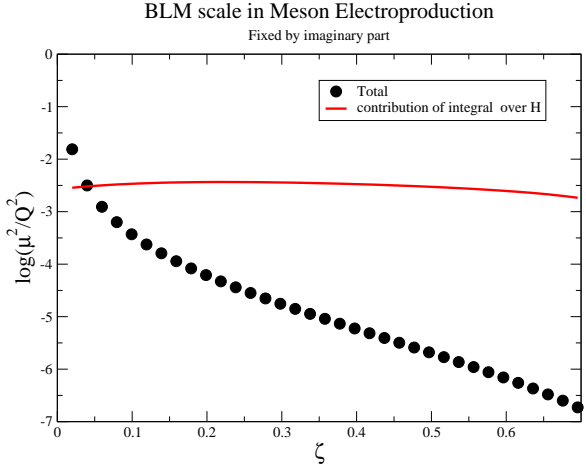,height=2.75in,angle=-90}
\caption[*]{\label{scaleplot} BLM scale in $\rho^0_L$ meson
electroproduction at $Q^2=4\ GeV^2$ and $t=-2\ GeV^2$. The solid
line represents the contribution of the integral in Eq.
(\ref{quotientscale}) }
\end{figure}

In Fig.~\ref{scaleplot} we display the resulting BLM scale as
extracted from the imaginary part of the amplitude. It can be seen
that $\mu_{BLM}^2 \ll Q^2$, indicating that meson electroproduction
at practically accessible kinematics measures the coupling constant $\alpha_s$
in the  infrared regime where theoretical considerations \cite{Alkofer:2004it}
lead us to expect a fixed point.
For numerical purposes one can use
\begin{equation} \label{alphas}
\alpha_s(\mu^2)=\frac{4\pi}{(9\log((\mu^2+M_0^2)/\Lambda^2))}
\end{equation}
with $\Lambda\simeq 0.2-0.21 \ GeV$ and $M_0 \simeq 1-1.1\ GeV$
at the BLM scale in the $\overline{MS}$ scheme.
With this, eqs. (\ref{immeson}, \ref{remeson}) can be immediately 
evaluated.
This legacy form of the running coupling has been successfully applied 
to the prediction of electromagnetic form factors and other quantities 
(see overview in \cite{Brodsky:1982nx}). Since the BLM scale that we 
determine is small, our calculations sample mostly $\alpha_s(0)$. 
The value of this fundamental quantity, controlled in eq. (\ref{alphas})
by the parameter $M_0$, is uncertain. With our parameterization,
$\alpha_s(0) \simeq 0.9$.

\begin{figure}
 \psfig{figure=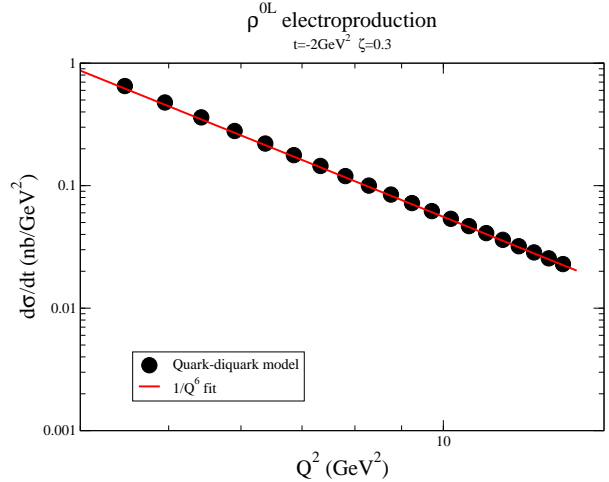,height=2.75in,angle=-90}
\caption{\label{rhoprod} As an illustration we present the model
computation for $\rho^0_L$ meson electroproduction at $t=-2\ GeV^2$
and $\zeta=0.3$. The fall-off with $Q^6$ at fixed $\zeta$ follows
from Eq. (\ref{rhoamp}) and (\ref{rhocross}). }
\end{figure}

As a final illustration we give the model's prediction for the
differential cross section in Eq. (\ref{rhocross}). The result of
the computation is plotted in Fig.~\ref{rhoprod}. Since the
perturbative diquark model is only viable at sizeable $t$, 
we fix $t=-2\ GeV^2$. Consequently the cross section is much smaller
than reported by other authors at $t=t_{\rm minimum}$. The fall-off
as a power law at fixed skewness is built-in into our formalism, and
can be traced to Eq. (\ref{rhoamp}), (\ref{rhocross}).  
(Notice that the DVCS cross section is enhanced at high momentum 
transfer by a factor $\frac{Q^2}{f_\rho^2}$ and eventually crosses 
over and dominates over $\rho^0$ production in spite of the extra 
$\alpha_{EM}$ suppression in Compton scattering \cite{inpreparation}). 
For this example we have chosen kinematics such that $Q^2/t$ is 2 or 
more as a compromise between theory and experimental uncertainties. 
From the theoretical point of view one would like to control the 
factorization corrections increasing this ratio. However a much larger
$Q^2$ would be difficult for an experiment at an upgraded Jefferson lab
or, due to statistical limitations at sizeable $t$, at larger $s$ machines.
Note that the largest uncertainty in our work likely comes from modeling 
the GPD H. 
The DVME $Q^{-6}$ fall-off is a genuine QCD prediction 
\cite{Brodsky:1994kf}
and consistent with the experimental result of HERMES
\cite{Airapetian:2000ni} at $\sqrt{s}=5.4 GeV$  that yields an
exponent $-5.66(36)$, within one standard deviation.
One should notice the subtlety that the Hermes data is quoted at fixed 
center of mass energy squared $W^2=s$ for the photon-nucleon system, 
whereas the predicted fall-off is at fixed scaling variable, here the 
skewness $\zeta$. But since reference \cite{Airapetian:2000ni} also 
displays a very weak $W$ dependence, one can slide $s$ in the Hermes 
data and use our eq.
(\ref{skewnessdef}) to obtain the same exponent at fixed skewness.
The  cross sections reported by this
experiment, which are dominated by low $t$ Regge processes, cannot be
directly compared with our result. We look forward to future
measurements with a larger $-t$ reach.

\section{Conclusion}

We have shown that  there is no {\it a priori}
difficulty to set the renormalization scales which appear in hard  
hadronic scattering amplitudes that are genuinely complex.   The 
individual BLM scales for 
the real and imaginary parts are unambiguous and distinct,  and they maintain the analyticity and 
dispersion relations of the full scattering amplitude.
Just as in Abelian theory, the BLM method in QCD fixes the arguments of the running coupling to maintain the correct cut structure of the quark loop vacuum polarization contribution at each flavor threshold.  All the non-conformal contributions to the amplitude which are associated with the running of the gauge coupling are used to set the renormalization scales.   The remaining perturbation series is then identical with a conformal theory with zero $\beta$ function.

We have determined  the BLM renormalization scales for the imaginary and real parts of 
the $\gamma^*  p \to \rho^0_L  ~ p^\prime$ meson electroproduction amplitude, a process of current
theoretical and experimental interest. We applied the BLM prescription 
to the imaginary part and employ a fixed-t dispersion relation to calculate
the scale-fixed real part.
We have exploited the
connection between generalized parton distributions  and the parton-proton scattering amplitude
in order to obtain an analytic representation of the helicity-conserving amplitude $H$ in terms of a
simple, yet physically inspired, quark-diquark model of the proton.

It has been conventional to characterize the precision of pQCD predictions by
using an arbitrary renormalization scale in the $\overline{MS}$ scheme,
such as $\mu^2_R = Q^2,$ and then varying the scale over an arbitrary
range, e.g., $Q^2/2  < \mu^2_R < 2~ Q^2$ as a means to estimate the
convergence of the  perturbative series.   However, the variation of
the renormalization scale can only be relevant for the nonconformal contributions to an observable, not the complete
series.  In fact, as we have stressed,  there is actually no renormalization scale ambiguity since one
must set the argument of the running coupling such that amplitudes have the correct
analytic cut structure at each quark threshold. This is done correctly
in any renormalization scheme by using the BLM method.

\begin{acknowledgement}
F.J. Llanes-Estrada warmly thanks the hospitality of the Stanford
Linear Accelerator Center's theory group where this work was carried
out, and partial financial support from a Fundacion del Amo-Univ.
Complutense fellowship. Work supported in part by grants FPA
2004-02602, 2005-02327, PR27/05-13955-BSCH (Spain) and U.S.
Department of Energy under contract number DE-AC02-76SF00515.
\end{acknowledgement}

\end{document}